\documentclass[
4pt,
superscriptaddress,
preprint,
 amsmath,amssymb,
 aps,
pra,
]{revtex4-1}

\usepackage{graphicx}
\usepackage{dcolumn}
\usepackage[colorlinks=true,
            linkcolor=blue,
            urlcolor=black,
            citecolor=blue]{hyperref}
\usepackage{bm}
\usepackage{enumerate}
\usepackage{lipsum}
\usepackage{threeparttable}
\usepackage{epstopdf}
\usepackage{float}
\usepackage{xcolor,colortbl}
\usepackage[caption = false]{subfig}
\usepackage{tabularx}

\begin{document}


\title{Dynamic polarizabilities and magic wavelengths of Sr$^+$ for focused vortex light}
\author{Anal Bhowmik}
\email{analbhowmik@phy.iitkgp.ernet.in}
\affiliation{Department of Physics, Indian Institute of Technology Kharagpur, Kharagpur-721302, India.}
\author{Sonjoy Majumder}
\email{sonjoym@phy.iitkgp.ernet.in}
\affiliation{Department of Physics, Indian Institute of Technology Kharagpur, Kharagpur-721302, India.}

\date{\today}





\begin{abstract}

A theory of dynamic polarizability for trapping relevant states of Sr$^+$ is presented here when the ions  interact with a focused optical vortex. The coupling between the  orbital and spin angular momentum of the optical vortex varies with focusing angle of the beam and  is studied in the calculation of  the magic wavelengths for  $5s_{{1}/{2}}\rightarrow 4d_{{3}/{2},  {5}/{2}}$ transitions of Sr$^+$. The initial state of our interest here is $5s_{{1}/{2}}$ with $m_J = -1/2$ of which is different possible trapping state compare to our recent work on Sr$^+$ [Phys. Rev. A \textbf{97}, 022511 (2018)]. We find  variation in magic wavelengths and the corresponding polarizabilities with different combinations of orbital and spin angular momentum of the  vortex beam. The variation is  very significant when the wavelengths of the beam are in the infrared region of electromagnetic spectrum.  The calculated magic wavelengths will help the  experimentalists to trap the ion for performing the high precision spectroscopic  measurements.

\end{abstract}

\maketitle
\section{INTRODUCTION}
  
Optical trapping of atoms or ions has been extensively used in  high precision spectroscopic measurements \cite{Champenois2004, Chou2010}.  But the mechanism of trapping  using a laser light inevitably produces a shift in the  energy levels of the atoms involved  in absorption.  The  shift is called the stark shift. In general,  the shift is  different for these energy states of  the atom. Thus naturally it  will influence the fidelity of the precision  measurement experiments due to non-achieving of exact resonance. However, this drawback can be diminished if the atoms are trapped at magic wavelengths of the laser beam, for which the  differential ac stark shift of an atomic transition effectively vanishes. Therefore, the magic wavelengths have significant applications in atomic clocks \cite{Margolis2009, Rosenbusch2009, Nicholson2015}, atomic
magnetometers \cite{Dong2015} and  atomic
interferometers \cite{Biedermann2015}.

 All the previous studies of magic wavelengths  for trapping of different atoms or ions are obtained for the Gaussian modes of a laser \cite{Flambaum2008, Arora2012, Arora2007, Ludlow2008}.  In this work, we determine the magic wavelengths of the transitions $5s_{1/2,-1/2} \rightarrow 4d_{3/2, m_J}$ 
 and $5s_{1/2,-1/2} \rightarrow 4d_{5/2, m_J}$ of Sr$^+$ ion, assuming the external light field is a circularly polarized focused optical vortex such as Laguerre-Gaussian (LG) beam \cite{Bhowmik2016}. Since the stark shifts will be different for  the states $5s_{1/2,-1/2}$ and $5s_{1/2,+1/2}$, different laser frequency (magic frequency) should be applied to minimize the systematic errors in the experiments involved the state $5s_{1/2,-1/2}$  compare to $5s_{1/2,+1/2}$ state. Therefore, it is important to quantify the magic wavelengths of the transitions $5s_{1/2,-1/2} \rightarrow 4d_{3/2, m_J}$ and  $5s_{1/2,-1/2} \rightarrow 4d_{5/2, m_J}$ of Sr$^+$, as we have already reported the magic wavelengths related to   $5s_{1/2,+1/2}$ state \cite{Bhowmik2018}. However, the special property of optical vortex is that, apart from the polarization (i.e., spin angular momentum (SAM)),  the optical vortex carries orbital angular momentum (OAM) due to its helical phase front \cite{Mondal2014}. Now, it is well known that during the interaction of a paraxial LG beam with atoms or ions (which
are below its recoil limit),  the quadrupole transition is the lowest-order transition where the OAM of the LG beam affects the electronic motion \cite{Mondal2014, Schmiegelow2016}. Therefore, the OAM of a paraxial LG beam does not influence dipole polarizability of an atomic state. Hence,  in case of paraxial LG beam,  the dipole polarizability and the magic wavelengths solely depend on the SAM of the beam. But unlike the paraxial LG beam,  the OAM and SAM of the optical vortex get coupled when the beam is focused \cite{Bhowmik2016}.  This leads to the transfer of OAM to the electronic motion of the atoms in the  dipole transition  level and creates an impact on the polarizability of an atomic state \cite{Bhowmik2016}. Further, the coupling of angular momenta increases with the focusing angle. However, in this work, we quantify all these effects of  OAM and SAM on the polarizability of an atomic state regarding magic wavelengths.

\section{THEORY}
 If an atom or ion placed in an external oscillating electric field $E(\omega)$, then the second-order shift in a particular energy level of the atom or ion is proportional to the square of the electric field,  $  
E^2(\omega)$. The proportional   coefficient is called  the dynamic polarizability $\alpha(\omega)$ of the atomic or ionic energy state at frequency $\omega$ of the external electric field and it can be written as
\cite{Dutta2015}
\begin{equation}\label{1}
\alpha(\omega)=\alpha_c(\omega)+\alpha_{vc}(\omega)+\alpha_v(\omega).
\end{equation}
 
Where $\alpha_c(\omega)$ and $\alpha_v(\omega)$ are dynamic  core polarizability of the ionic core  and dynamic valence polarizability of the single valence system, respectively.  This ionic
core is obtained by removing the valence electron from the
system. $ \alpha_{vc}(\omega) $ is the correction \cite{Safronova2011} in core polarizability in the presence of the valence electron.  
As the core electrons are tightly bound to
the nucleus, the presence of a valence electron is expected not to
change the core polarizability significantly. Thus we
consider $ \alpha_{vc}$  in the present method of calculations without variation of $\omega$. 
$\alpha_v(\omega) $ is calculated using the external electric field of focused LG beam \cite{Bhowmik2016}. In case of focused LG beam, OAM and SAM are no longer separately a good quantum number as they get coupled to each other. Therefore, the effect of total angular momentum (OAM+SAM) can be seen on  $\alpha_v(\omega) $, which can be expressed  as \cite{Bhowmik2018}

\begin{eqnarray}\label{2}
\alpha_v(\omega) =2A_0 \alpha_v^0(\omega)
+ 2\times\left(\frac{m_J}{2J_v}\right) A_1\alpha_v^1(\omega)  
&+& 2 \times \left(\frac{3m_J^2-J_v(J_v+1)}{2J_v(2J_v-1)}\right)A_2\alpha_v^2(\omega),
\end{eqnarray}
where $J_v$ is the total angular momentum of the state $\psi_v$ and $m_J$  is its  magnetic component. The coefficients $A_i$s are  
 $A_0=\left[\{I_0^{(l)}\}^2+\{I_{\pm 2}^{(l)}\}^2+2 \{I_{\pm 1}^{(l)}\}^2\right]$,
 $A_1=\left[\pm \{I_0^{(l)}\}^2 \mp \{I_{\pm 2}^{(l)}\}^2\right]$
 and $A_2= \left[\{I_0^{(l)}\}^2+\{I_{\pm 2}^{(l)}\}^2-2 \{I_{\pm 1}^{(l)}\}^2\right]$. The parameter $I_m^{(l)}$, where $m$ takes the values 0, $\pm1$ and $\pm2$,    depends on focusing angle ($\theta_{max}$) by \cite{Zhao2007, Bhowmik2016}

\begin{equation}\label{3}
I_m^{(l)}(r _\bot ^\prime ,z ^\prime)=\int_0^{\theta_{max}}d\theta\left({\dfrac{\sqrt{2}r_\bot^\prime }{w_0 \sin\theta}}\right)^{\lvert l \rvert}{(\sin\theta)}^{\lvert l \rvert +1} 
\sqrt{\cos\theta}g_{\lvert m \rvert}(\theta) J_{l+m}(kr_\bot^\prime \sin\theta)e^{ikz^\prime \cos\theta}.
\end{equation} 
Here $r_\bot^\prime$ is the projection of \textbf{r$^\prime$} on the $xy$ plane, $w_0$ is the waist of the paraxial circularly polarized LG beam which is focused by a high numerical aperture. The angular functions are  $g_0 (\theta)=1+\cos\theta$, $g_1 (\theta)=\sin\theta$ and $g_2 (\theta)=1-\cos\theta$.   $\alpha_v^0(\omega)$, $ \alpha_v^1(\omega) $ and $ \alpha_v^2(\omega)$ introduced in Eq.(~\ref{2}) are the scalar, vector and tensor parts  respectively, of the valence polarizability and are expressed as \cite{Mitroy2010, Dutta2015} 
 \begin{equation}\label{7}
 \alpha_v^0(\omega)=\frac{2}{3(2J_v+1)}\sum_n \frac{|\langle\psi_v||d||\psi_n\rangle|^2\times(\epsilon_n-\epsilon_v)}{(\epsilon_n-\epsilon_v)^2-\omega^2},
 \end{equation}
\begin{equation}\label{8}
\alpha_v^1(\omega)=-\sqrt{\frac{6J_v}{(J_v+1)(2J_v+1)}}\sum_n (-1)^{J_n+J_v} \left\{\begin{array}{ccc} J_v & 1 & J_v\\ 1 & J_n& 1 \end{array}\right \}\frac{|\langle\psi_v||d||\psi_n\rangle|^2 \times 2\omega}{(\epsilon_n-\epsilon_v)^2-\omega^2},
\end{equation}
and
\begin{equation}\label{9}
\alpha_v^2(\omega)=4\sqrt{\frac{5J_v(2J_v-1)}{6(J_v+1)(2J_v+1)(2J_v+3)}}\sum_n (-1)^{J_n+J_v} \left\{\begin{array}{ccc} J_v & 1 & J_n\\ 1 & J_v& 2 \end{array}\right \} \frac{|\langle\psi_v||d||\psi_n\rangle|^2\times(\epsilon_n-\epsilon_v)}{(\epsilon_n-\epsilon_v)^2-\omega^2}.
\end{equation}
Henceforth, whenever we mention about SAM or OAM in the following text, it is considered to be the angular momentum of the paraxial LG beam before passing through the focusing lens.

\section{NUMERICAL RESULTS AND DISCUSSIONS}
The aim of this work is to calculate  the dynamic polarizabilities of the $5s_{1/2}$, $4d_{3/2}$, and $4d_{5/2}$ states  for different magnetic sublevels of Sr$^+$. The scalar, vector and tensor parts of the valence polarizabilities are calculated using Eqs ~(\ref{7}), ~(\ref{8}), and ~(\ref{9}). The precise estimations of these three parts of the valence polarizability depend on the accuracy of the unperturbed energy levels and the dipole matrices among them. In order to evaluate these properties, we use correlation exhaustive relativistic coupled cluster (RCC) theory \cite{Dutta2016, Bhowmik2017a, Bhowmik2017b, Das2018, Biswas2018} with wave operators associated with single and double and partial triple excitations in linear and non-linear forms. The wavefunctions calculated by the RCC method can  produce highly precise \textit{E1} transition amplitudes  as discussed in our recent work \cite{Bhowmik2018}. Calculation in this reference yields that the static core polarizability ($\alpha_c(0)$) of the ion is 6.103 a.u., and the  static core-valence parts of the polarizabilities ($\alpha_{vc}(0)$) for the states $5s_{\frac{1}{2}}$, $4d_{\frac{3}{2}}$ and  $4d_{\frac{5}{2}}$ are  $-0.25$ a.u., $-0.38$ a.u. and $-0.42$ a.u., respectively.

In order to determine the precise values of dynamic valence polarizabilities, we require  calculating a large number of dipole matrix elements. Another way to say, the running index $n$ in  Eqs ~(\ref{7}) to ~(\ref{9}) is turning out to be around  25 for Sr$^+$ to obtain  accurate valence polarizability. Since the RCC method is  computationally very expensive, we break our total calculations of valence polarizability in three parts depending on their significance in the sums of Eqs ~(\ref{7})--(\ref{9}).   The first part includes the most  important contributing terms to the valence polarizabilities which involves  the $E1$ matrix elements associated  with the intermediate states  from $5^2P$ to $8^2P$ and $4^2F$ to $6^2F$. Therefore, these matrix elements  are calculated using the correlation exhaustive RCC method. The second part consists of the comparatively less significant terms associated with $E1$ matrix elements in the polarizability expressions arising from  intermediate states from  $9^2P$ to $12^2P$  and $7^2F$ to $12^2F$. Thus we calculate the second part using second-order relativistic many-body perturbation theory \cite{Johnson1996}. The last part, whose contributions are comparatively further small  to the valence polarizability,  includes the intermediate states from $ n=13$ to $ 25$, are computed using the Dirac Fock  wavefunctions.

In FIG.  1 and 2, we present    the variations of total polarizabilities of $5s_{1/2,-1/2}$, $4d_{3/2,m_J}$ and $4d_{5/2,m_J}$ (for different  magnetic quantum number, $m_j$,  of the states)   states with the frequency of   the external field of  the focused LG beam. The focusing angle of the LG beam is considered  50$^\circ$ in both the figures.  The combinations of  angular momenta of the paraxial LG  beam  have chosen as (OAM, SAM) $= (+1, +1)$ and $(+1, -1)$ in Fig. 1 and Fig. 2, respectively. The resonances occur in the plots due to   the $5s_{1/2}  \rightarrow 5p_{1/2, 3/2} $ transitions for $5s_{1/2}$ state, $4d_{3/2}  \rightarrow 5p_{1/2, 3/2} $ transitions for $4d_{3/2}$ state and $4d_{5/2}  \rightarrow 5p_{ 3/2}  $ transitions for $4d_{5/2}$ state. The plots show a number of intersections between the polarizabilities of  $5s_{\frac{1}{2}}$ at $m_J=-1/2$ and different multiplets of $4d_{\frac{3}{2}, \frac{5}{2}}$ states. These intersections indicate magic wavelengths, at which the difference in the stark shifts of the two related states vanishes. Figures show that magic wavelengths which fall in the infrared region of the electromagnetic spectrum have large  polarizabilities compared to the magic wavelengths of the visible or ultraviolet region. These magic wavelengths with high polarizabilities will be more effective   to trap the ion, and thus they are highly recommended for trapping. These two figures are given as an example.  Similar plots are studied for different focusing angles, say 60$^\circ$ and 70$^\circ$, and corresponding magic wavelengths are discussed later in this paper.

In Table I, II and III, we have listed a large number of magic wavelengths along with their corresponding polarizabilities, when the focusing angles of LG beam are 50$^\circ$, 60$^\circ$ and 70$^\circ$. The table I is for the transition $5s_{{1}/{2}}\rightarrow 4d_{{3}/{2}}$,  and  the combinations of OAM and SAM are  $(+1, +1)$, $(+1, -1)$, $(+2, +1)$ and $(+2, -1)$. In Table II and III, the transition is  $5s_{{1}/{2}}\rightarrow 4d_{{5}/{2}}$ but the combinations of OAM and SAM are $((+1, +1)$, $(+1, -1))$  and $((+2, +1)$,  $(+2, -1))$, respectively. The $m_J$ value of $5s_{{1}/{2}}$ is considered $-1/2$ throughout this paper and the tables show totally distinct set of magic wavelengths compared to the results published \cite{Bhowmik2018} considering $m_J=1/2$. There are five sets of magic wavelengths obtained for each of the multiplets  of $4d_{{3}/{2}}$ for all the combinations of angular momenta and focusing angles of the LG beam  in the given frequency range. Whereas, in the same range of wavelength spectrum for $4d_{{5}/{2}}$ state, our calculations show  seven sets of magic wavelengths  (see Table II and III) for most of the multiplets.   Since the resonance transition of $5s_{\frac{1}{2}} \rightarrow 4d_{\frac{3}{2}}, 4d_{\frac{5}{2}}$ are 687 nm and 674 nm, respectively, thus the ion is attracted and trapped  to the high intensity (low intensity) region of the LG beam when the  magic wavelength is larger (smaller)  than the resonance wavelength.

Since this work is about  the finding of suitable magic wavelengths for trapping,  we only give an estimation of the theoretical uncertainty in the calculated magic wavelengths. Here we collect the most important set of $E1$ matrix elements which include $5s_{\frac{1}{2}} \rightarrow 5p_{\frac{1}{2}, \frac{3}{2}}$ transitions for $5s_{\frac{1}{2}}$ state; $4d_{\frac{3}{2}} \rightarrow 5p_{\frac{1}{2}, \frac{3}{2}}$ and $4d_{\frac{3}{2}} \rightarrow 4f_{\frac{5}{2}}$  transitions for $4d_{\frac{3}{2}}$ state; $4d_{\frac{5}{2}} \rightarrow 5p_{\frac{3}{2}}$ and $4d_{\frac{5}{2}} \rightarrow 4f_{\frac{5}{2}, \frac{7}{2}}$ transitions for $4d_{\frac{5}{2}}$ state. We compare our RCC results with  the  SDpT values calculated by Safronova \cite{Safronova2010} and further apply those  $E1$ matrix elements in place of  our present RCC values to recalculate the magic wavelengths. This approach leads to the theoretical uncertainty in our calculated magic wavelength values is about $\pm$1\%.

\section{CONCLUSIONS}

In conclusions, we find a  wide list of magic wavelengths for the transitions $5s_{1/2}(m_J=-1/2) $  $\rightarrow$ $4d_{3/2}(m_J) $ and $5s_{1/2}(-1/2)  $  $\rightarrow$ $4d_{5/2}(m_J)$  of the Sr$^+$ ion. We have found here a quite  distinct set of values of magic wavelengths compared to the same with the initial state  $5s_{1/2}(m_J=1/2) $.  These magic wavelengths fall  from infrared to vacuum-ultraviolet range in the electromagnetic spectrum and will help experimentalists to trap ions in either at the high intensity or  the low intensity region of the LG beam. The variations in the magic wavelength are found tunable by varying  the  OAM, SAM and focusing angles of the LG beam. An appreciable amount of deviations concerning focusing angles in the infrared   magic wavelengths and  corresponding polarizabilities are observed. As these infrared magic wavelengths have significant large values of polarizabilities, they are recommended as the best for trapping in the high precision experiments.

\begin{figure*}[!h]
\subfloat[]{\includegraphics[trim = 1cm 1.0cm 0.1cm 1.5cm, scale=.40]{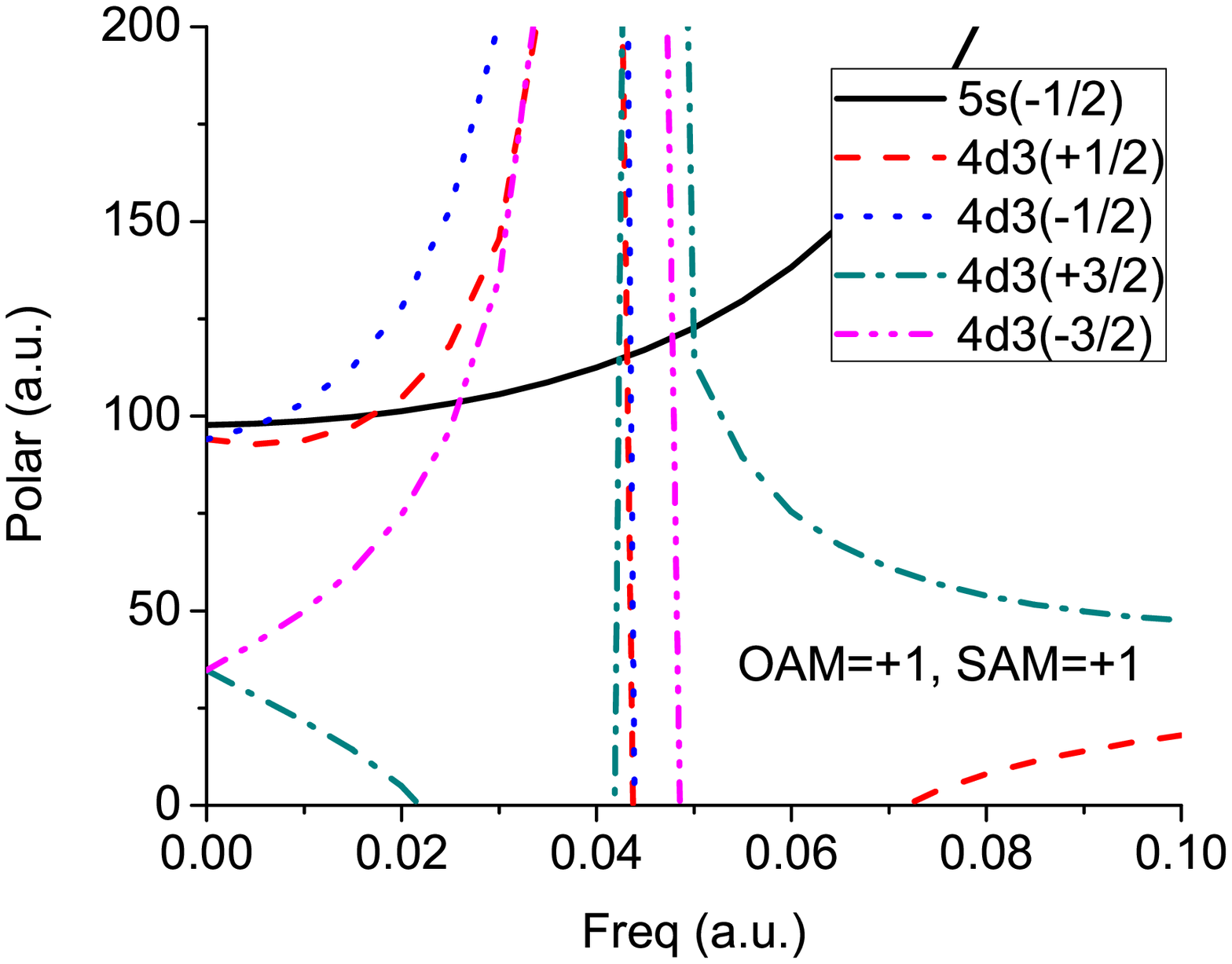}}
\subfloat[]{\includegraphics[trim = 1cm 1.0cm 0.1cm 1.5cm, scale=.40]{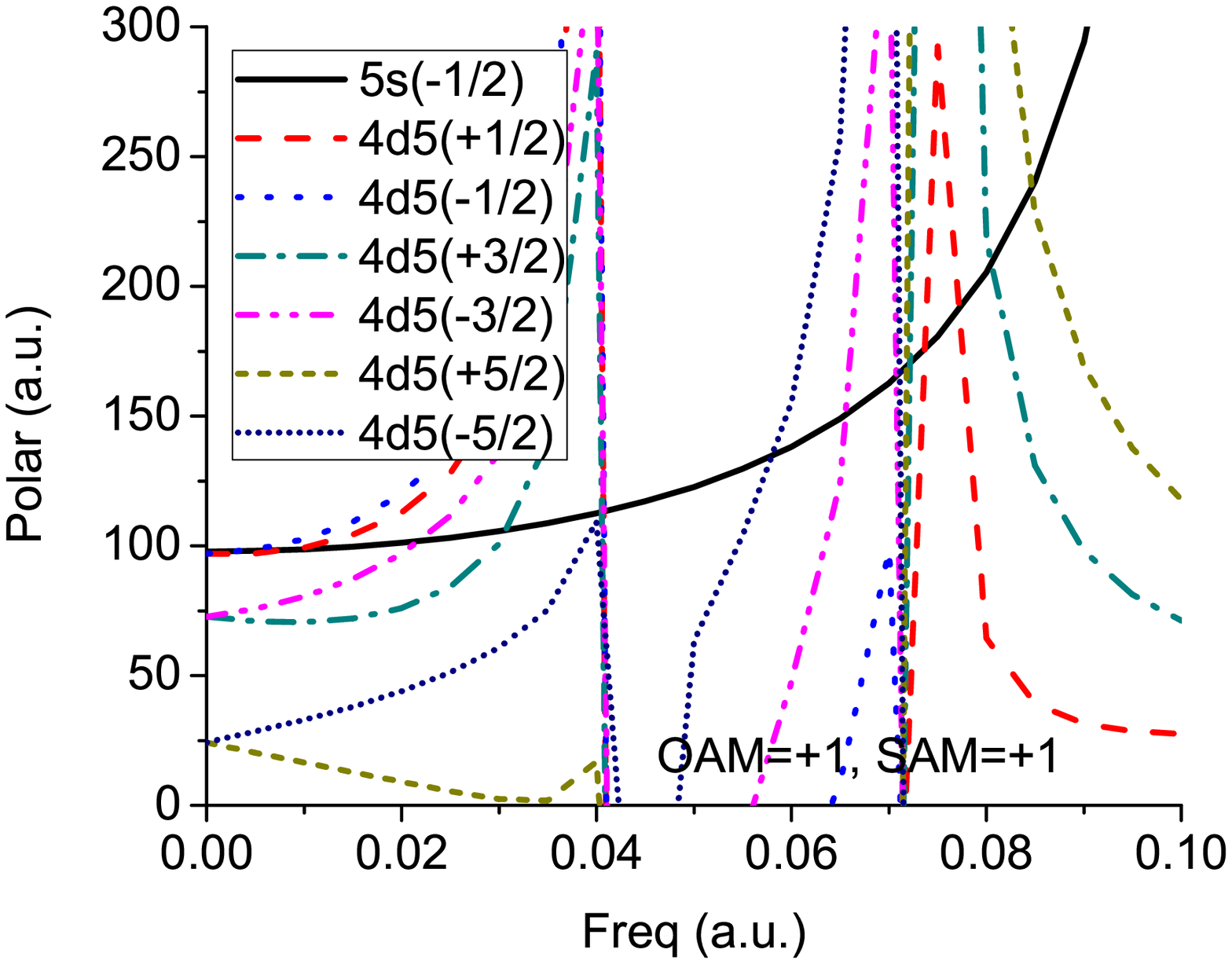}}\\
\subfloat[]{\includegraphics[trim =  1cm 1.0cm 0.1cm 0.1cm,scale=.40]{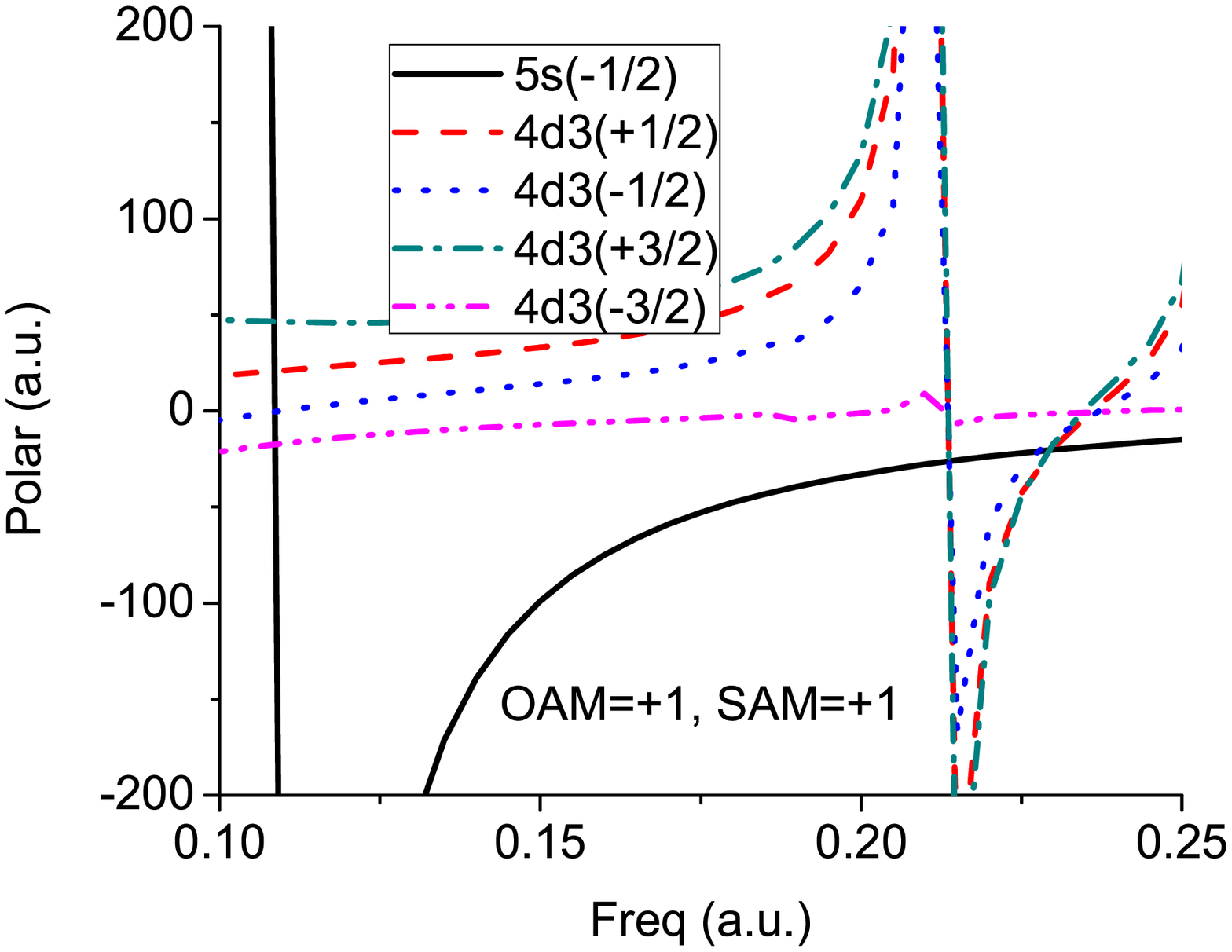}}
\subfloat[]{\includegraphics[trim =  1cm 1.0cm 0.1cm 0.1cm, scale=.40]{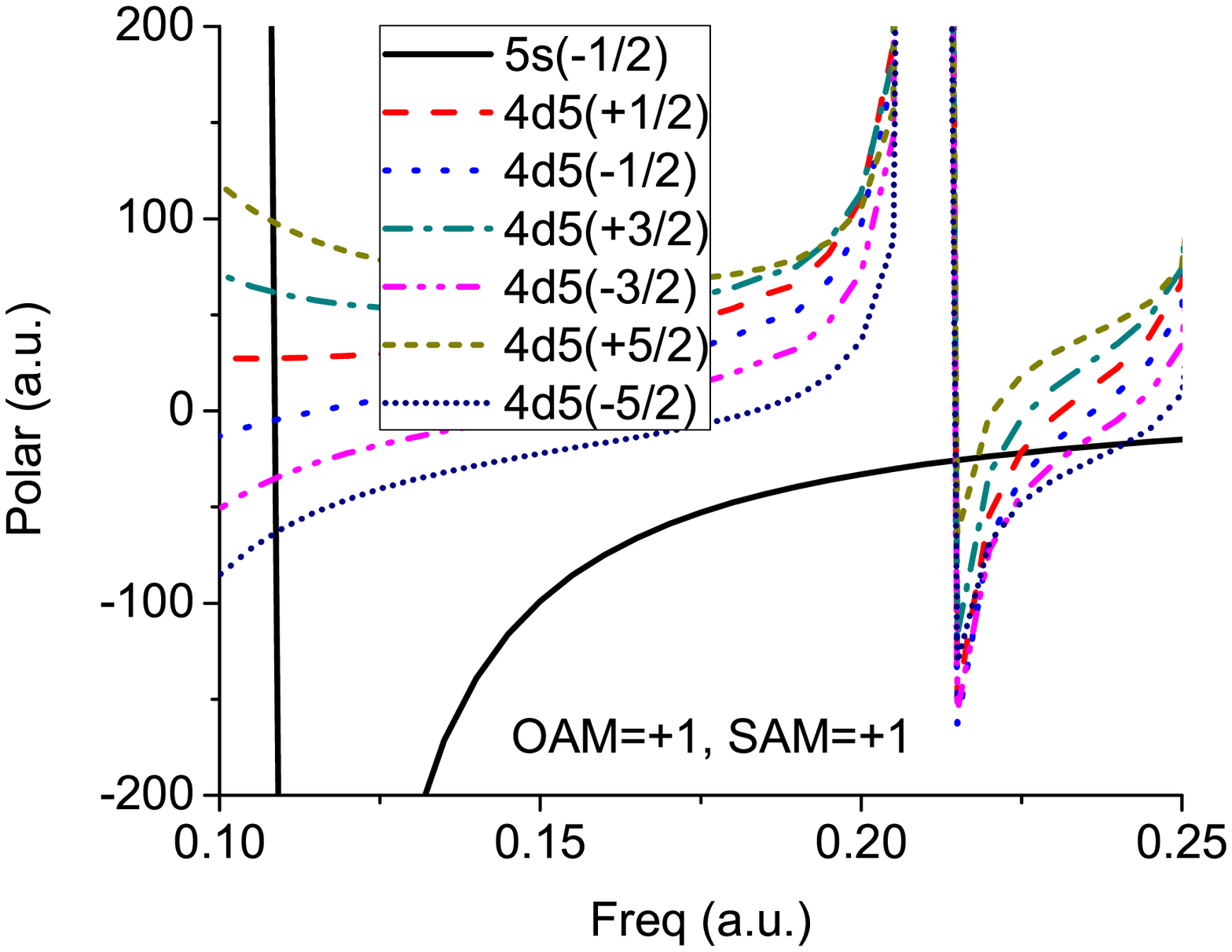}}
\caption{Variation of polarizabilities (Polar) of  $5s_{\frac{1}{2}}$ and $4d_{\frac{3}{2},\frac{5}{2}}$ states with frequency (Freq) are plotted when the focusing angle of LG beam is 50$^\circ$  with OAM=+1 and SAM=+1. The brackets indicate the magnitudes of different magnetic components. Fig. (a) and (c) are for the $5s_{\frac{1}{2}}$ and $4d_{\frac{3}{2}}$ states, and  Fig. (b) and (d) are for the $5s_{\frac{1}{2}}$ and $4d_{\frac{5}{2}}$ states.}
\end{figure*}

\begin{figure*}[!h]
\subfloat[]{\includegraphics[trim = 1cm 1.0cm 0.1cm 1.5cm, scale=.40]{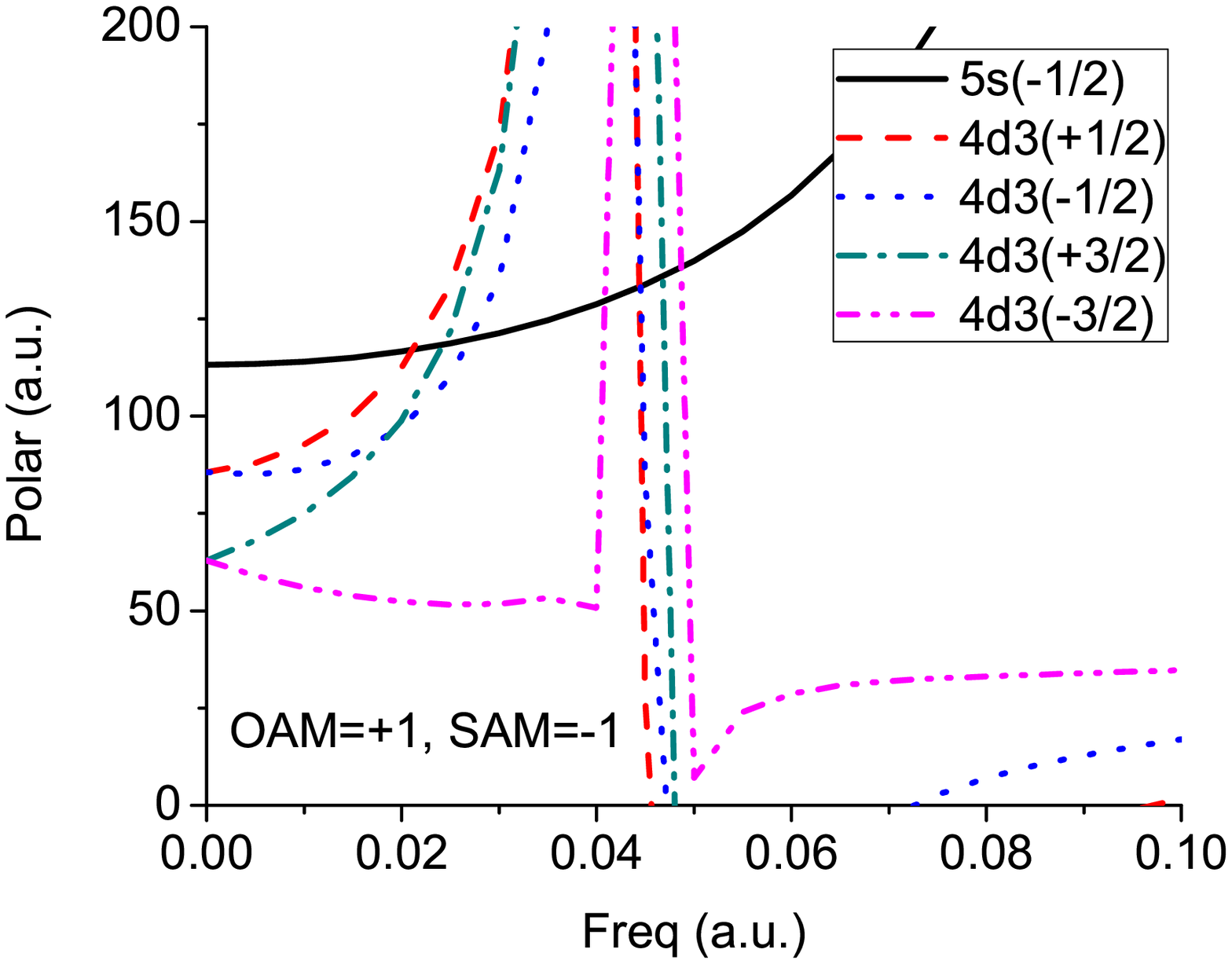}}
\subfloat[]{\includegraphics[trim = 1cm 1.0cm 0.1cm 1.5cm, scale=.40]{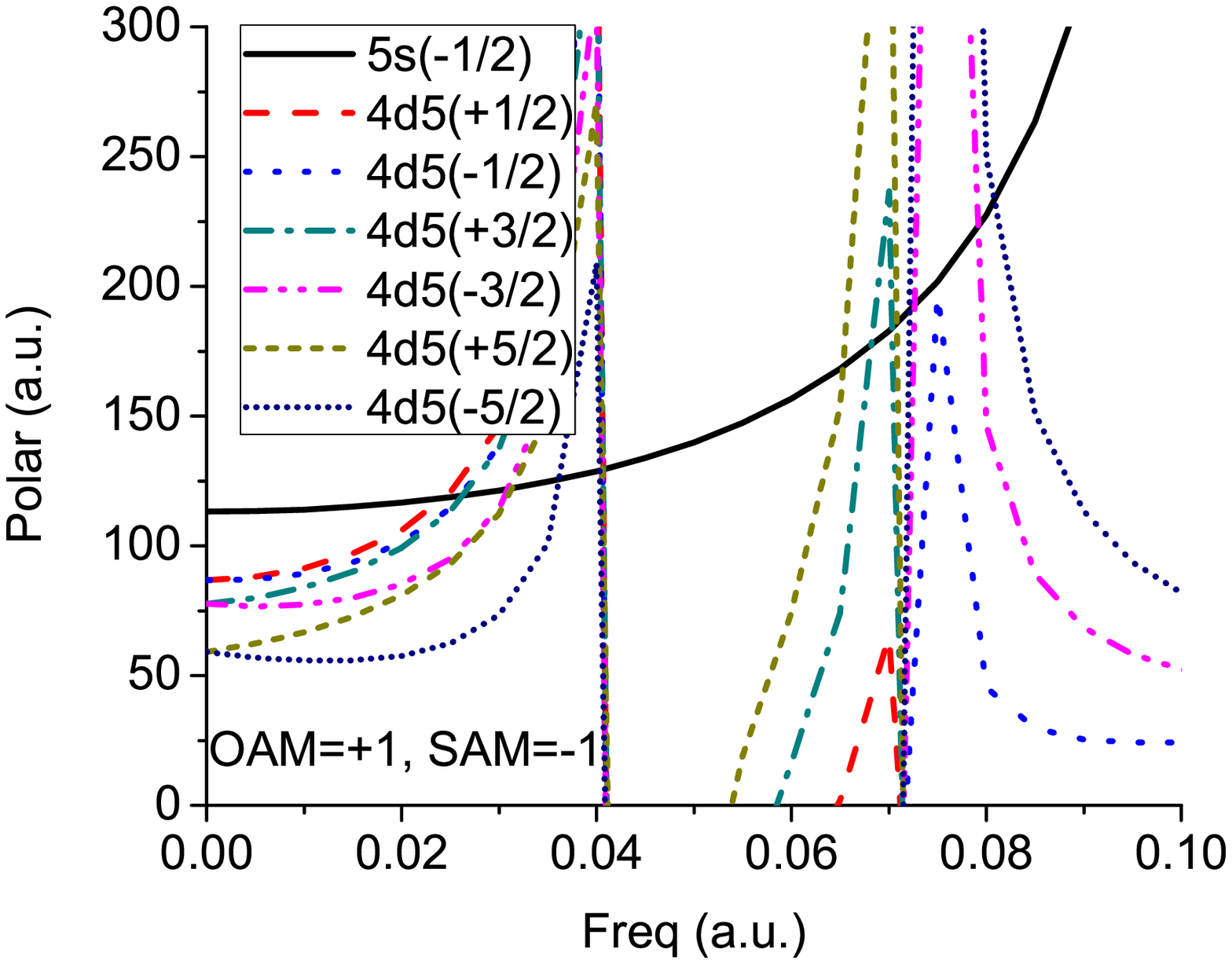}}\\
\subfloat[]{\includegraphics[trim =  1cm 1.0cm 0.1cm 0.1cm,scale=.40]{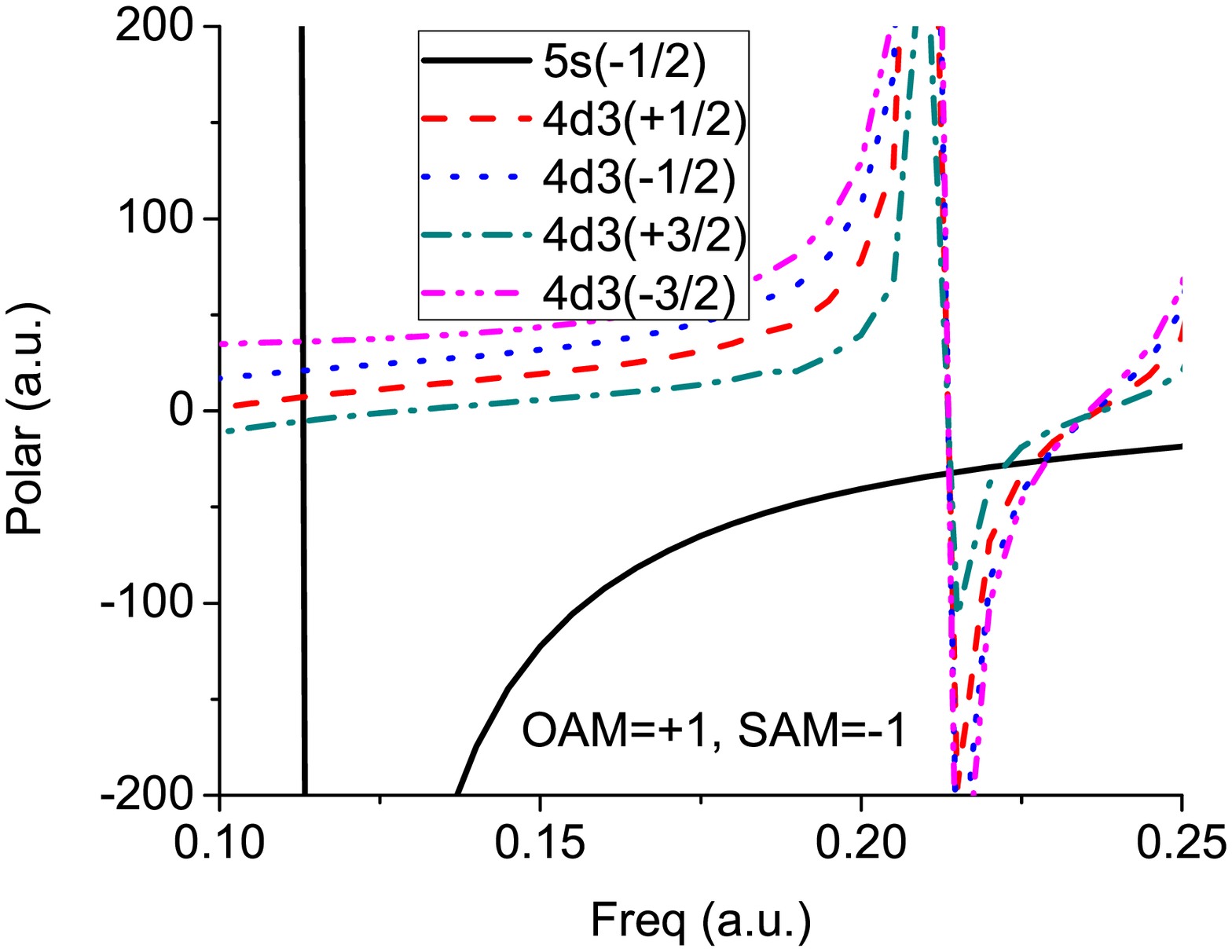}}
\subfloat[]{\includegraphics[trim =  1cm 1.0cm 0.1cm 0.1cm, scale=.40]{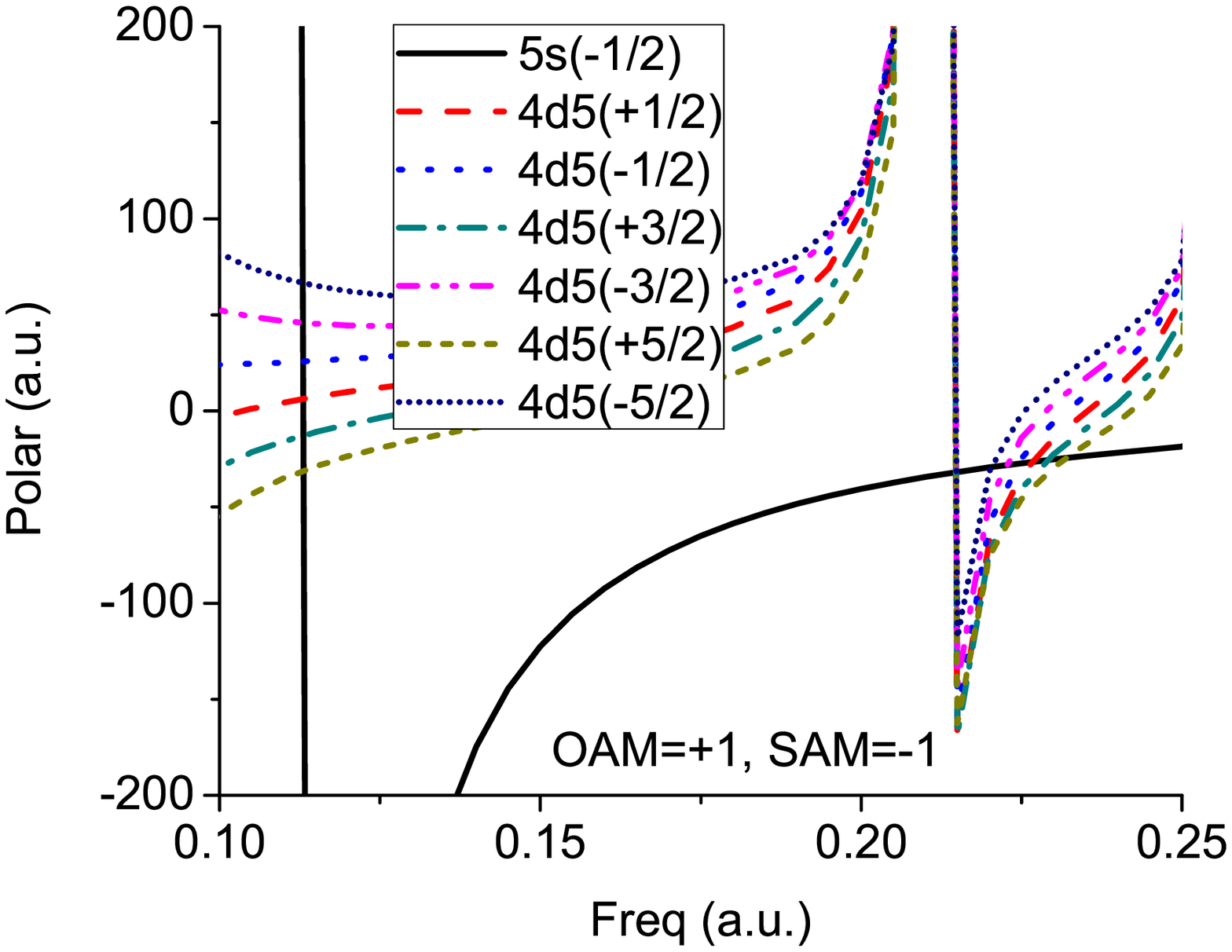}}
\caption{Variation of polarizabilities (Polar) of  $5s_{\frac{1}{2}}$ and $4d_{\frac{3}{2},\frac{5}{2}}$ states with frequency (Freq) are plotted when the focusing angle of LG beam is 50$^\circ$  with OAM=+1 and SAM=$-1$. The brackets indicate the magnitudes of different magnetic components.   Fig. (a) and (c) are for the $5s_{\frac{1}{2}}$ and $4d_{\frac{3}{2}}$ states, and  Fig. (b) and (d) are for the $5s_{\frac{1}{2}}$ and $4d_{\frac{5}{2}}$ states.}
\end{figure*}

\begin{table}[h]
\scriptsize
  \caption{Magic wavelengths (in nm)  of Sr$^+$ for  focusing angles 50$^\circ$, 60$^\circ$ and 70$^\circ$ of the LG beam for the transitions $5s_{1/2}(-1/2)  $  $\rightarrow$ $4d_{3/2}(m_J)  $.}
\centering
\begin{tabular}{cccccccccccccc}

\hline \hline
     \multicolumn{14}{c}{\textbf{Non-paraxial LG beam}}\\ 

   State ($4d_{3/2}(m_J)$)     &$\lambda_{\textrm{magic}}^{50^\circ}$&$\alpha$ & $\lambda_{\textrm{magic}}^{60^\circ}$&$\alpha$&  $\lambda_{\textrm{magic}}^{70^\circ}$&$\alpha$& State     ($4d_{3/2}(m_J)$) & $\lambda_{\textrm{magic}}^{50^\circ}$&$\alpha$ & $\lambda_{\textrm{magic}}^{60^\circ}$&$\alpha$&  $\lambda_{\textrm{magic}}^{70^\circ}$&$\alpha$
           \\ [0.2ex]
   \hline 
 & \multicolumn{6}{c}{\textbf{OAM=+1, SAM=+1}} && \multicolumn{6}{c}{\textbf{OAM=+1, SAM=-1}}\\  
  \hline  
$(+1/2)$	&	2680.20	&	100.36	&	2201.13	&	104.81	&	1963.94	&	109.41	&	$(+1/2)$	&	2169.68	&	117.01	&	2080.52	&	118.77	&	2016.08	&	120.32	\\
	&	1069.56	&	114.82	&	1054.71	&	119.84	&	1047.43	&	124.19	&		&	1037.89	&	132.76	&	1023.90	&	134.76	&	1021.60	&	136.67	\\
	&	422.27	&	20.89	&	421.10	&	20.50	&	417.25	&	20.13	&		&	404.65	&	7.19	&	404.65	&	7.69	&	404.65	&	8.47	\\
	&	213.61	&	-26.26	&	214.31	&	-27.87	&	213.51	&	-28.82	&		&	213.21	&	-32.25	&	213.11	&	-32.87	&	214.01	&	-33.72	\\
	&	198.27	&	-20.34	&	198.45	&	-21.32	&	198.71	&	-22.37	&		&	200.72	&	-26.41	&	200.72	&	-26.69	&	200.81	&	-27.23	\\
$(-1/2)$	&	8136.31	&	98.17	&	4339.37	&	102.62	&	3120.78	&	106.42	&	$(-1/2)$	&	1693.80	&	119.80	&	1668.99	&	121.56	&	1650.85	&	122.76	\\
	&	1069.56	&	114.98	&	1042.64	&	119.56	&	1040.26	&	124.45	&		&	1021.60	&	133.40	&	1019.31	&	135.28	&	1014.77	&	137.07	\\
	&	421.88	&	-0.84	&	420.71	&	0.33	&	416.87	&	2.29	&		&	404.65	&	20.87	&	404.65	&	20.76	&	404.65	&	20.22	\\
	&	213.11	&	-26.13	&	214.11	&	-27.80	&	213.21	&	-28.51	&		&	213.21	&	-32.25	&	213.11	&	-32.87	&	214.31	&	-33.80	\\
	&	200.19	&	-20.96	&	200.28	&	-22.04	&	200.37	&	-23.13	&		&	199.31	&	-25.72	&	199.40	&	-26.33	&	199.49	&	-26.61	\\
$(+3/2)$	&	1074.61	&	114.75	&	1079.70	&	117.94	&	1082.26	&	122.18	&	$(+3/2)$	&	1875.04	&	118.41	&	1875.04	&	119.83	&	1875.04	&	121.14	\\
	&	911.27	&	122.62	&	916.77	&	126.69	&	920.47	&	130.40	&		&	971.50	&	136.00	&	977.75	&	137.55	&	977.75	&	138.96	\\
	&	422.67	&	46.61	&	421.49	&	44.61	&	417.25	&	42.08	&		&	404.29	&	-5.33	&	404.29	&	-4.11	&	404.29	&	-2.46	\\
	&	213.91	&	-26.39	&	214.31	&	-27.87	&	213.61	&	-28.82	&		&	212.91	&	-31.91	&	212.91	&	-32.53	&	213.31	&	-33.37	\\
	&	198.71	&	-20.47	&	198.71	&	-21.40	&	198.71	&	-22.51	&		&	204.78	&	-28.38	&	204.23	&	-28.50	&	203.77	&	-28.53	\\
$(-3/2)$	&	1759.20	&	103.64	&	1786.80	&	106.72	&	1800.92	&	110.43	&	$(-3/2)$	&	1114.02	&	129.69	&	1130.60	&	130.60	&	1199.04	&	130.05	\\
	&	957.21	&	120.05	&	957.21	&	124.22	&	963.28	&	127.90	&		&	935.59	&	138.21	&	939.45	&	139.68	&	945.30	&	141.12	\\
	&	421.49	&	-17.31	&	420.71	&	-15.76	&	416.87	&	-12.98	&		&	405.01	&	35.93	&	405.01	&	34.62	&	405.01	&	33.22	\\
	&		&		&		&		&	212.52	&	-28.35	&		&	213.21	&	-32.25	&	213.11	&	-32.87	&	214.31	&	-33.80	\\
	&		&		&		&		&	208.53	&	-26.52	&		&	198.97	&	-25.49	&	199.05	&	-25.99	&	199.14	&	-26.42	\\

 \hline 
 & \multicolumn{6}{c}{\textbf{OAM=+2, SAM=+1}} && \multicolumn{6}{c}{\textbf{OAM=+2, SAM=-1}}\\  
  \hline

$(+1/2)$	&	2462.88	&	101.83	&	2052.40	&	107.37	&	1822.53	&	113.66	&	$(+1/2)$	&	2149.21	&	117.80	&	2052.40	&	119.38	&	1955.51	&	121.23	\\
	&	1057.15	&	116.84	&	1052.27	&	122.12	&	1037.89	&	127.22	&		&	1021.60	&	134.04	&	1023.90	&	136.07	&	1019.31	&	137.47	\\
	&	419.55	&	20.60	&	418.01	&	20.67	&	415.72	&	21.18	&		&	404.65	&	7.20	&	404.65	&	7.77	&	404.65	&	8.96	\\
	&	213.21	&	-26.96	&	213.21	&	-27.86	&	213.11	&	-29.72	&		&	213.81	&	-32.84	&	213.31	&	-33.02	&	213.81	&	-33.82	\\
	&	198.36	&	-20.70	&	1986.20	&	-21.98	&	198.88	&	-23.52	&		&	200.81	&	-26.65	&	200.72	&	-26.97	&	200.81	&	-27.44	\\
$(-1/2)$	&	6417.37	&	100.05	&	3704.34	&	104.54	&	2696.06	&	110.28	&	$(-1/2)$	&	1687.53	&	119.86	&	1662.90	&	121.92	&	1633.10	&	123.97	\\
	&	1047.43	&	117.88	&	1040.26	&	122.15	&	1037.89	&	127.22	&		&	1021.60	&	134.04	&	1019.31	&	136.45	&	1008.04	&	138.16	\\
	&	419.17	&	0.34	&	417.63	&	1.67	&	415.35	&	3.45	&		&	405.01	&	20.78	&	405.01	&	20.42	&	405.01	&	20.08	\\
	&	213.21	&	-26.96	&	213.21	&	-27.86	&	213.11	&	-29.72	&		&	214.21	&	-33.10	&	213.71	&	-33.31	&	214.11	&	-34.02	\\
	&	200.19	&	-21.41	&	200.37	&	-22.68	&	200.54	&	-24.18	&		&	199.31	&	-25.96	&	199.49	&	-26.46	&	199.58	&	-26.88	\\
$(+3/2)$	&	1077.15	&	116.11	&	1082.26	&	119.92	&	1090.03	&	125.50	&	$(+3/2)$	&	1875.04	&	118.48	&	1875.04	&	120.23	&	1859.73	&	121.61	\\
	&	913.09	&	123.69	&	918.62	&	129.07	&	924.21	&	133.98	&		&	975.66	&	136.02	&	977.75	&	138.62	&	979.86	&	139.45	\\
	&	419.94	&	45.49	&	418.40	&	43.48	&	415.72	&	39.44	&		&	404.29	&	-4.96	&	404.65	&	-3.42	&	404.65	&	-1.53	\\
	&	213.21	&	-26.96	&	213.21	&	-27.86	&	213.11	&	-29.72	&		&	213.01	&	-32.49	&	213.01	&	-32.97	&	213.21	&	-33.53	\\
	&	198.71	&	-20.82	&	198.71	&	-22.04	&	198.79	&	-23.48	&		&	204.60	&	-28.39	&	203.95	&	-28.51	&	203.50	&	-28.66	\\
$(-3/2)$	&	1772.89	&	105.38	&	1793.83	&	109.29	&	1822.53	&	113.66	&	$(-3/2)$	&	1119.49	&	130.00	&	1139.08	&	131.07	&	1238.13	&	130.00	\\
	&	955.21	&	122.16	&	961.25	&	125.59	&	967.37	&	131.68	&		&	937.52	&	138.77	&	941.39	&	140.31	&	947.26	&	141.51	\\
	&	419.17	&	-16.99	&	417.63	&	-14.64	&	415.35	&	-11.16	&		&	405.01	&	35.64	&	405.01	&	33.98	&	405.01	&	32.12	\\
	&		&		&	212.22	&	-27.86	&	212.71	&	-29.20	&		&	214.31	&	-33.18	&	213.81	&	-33.42	&	214.21	&	-34.17	\\
	&		&		&	209.78	&	-26.52	&	207.48	&	-26.86	&		&	199.05	&	-25.81	&	199.05	&	-26.17	&	199.14	&	-26.63	\\

\hline
\hline
\label{table:nonlin} 
\label{I}

\end{tabular}
\end{table}

\begin{table}[h]
\scriptsize
  \caption{Magic wavelengths (in nm)  of Sr$^+$ for different focusing angles 50$^\circ$, 60$^\circ$ and 70$^\circ$ of the LG beam for the transitions $5s_{1/2}(-1/2)  $  $\rightarrow$ $4d_{5/2}(m_J)$.}
\centering
\begin{tabular}{cccccccccccccc}

\hline \hline
     \multicolumn{14}{c}{\textbf{Non-paraxial LG beam}}\\ 

   State ($4d_{5/2}(m_J)$)     & $\lambda_{\textrm{magic}}^{50^\circ}$&$\alpha$ & $\lambda_{\textrm{magic}}^{60^\circ}$&$\alpha$&  $\lambda_{\textrm{magic}}^{70^\circ}$&$\alpha$& State     ($4d_{5/2}(m_J)$) & $\lambda_{\textrm{magic}}^{50^\circ}$&$\alpha$ & $\lambda_{\textrm{magic}}^{60^\circ}$&$\alpha$&  $\lambda_{\textrm{magic}}^{70^\circ}$&$\alpha$
           \\ [0.2ex]
   \hline 
 & \multicolumn{6}{c}{\textbf{OAM=+1, SAM=+1}} && \multicolumn{6}{c}{\textbf{OAM=+1, SAM=-1}}\\  
  \hline  
$(+1/2)$	&	5841.46	&	98.42	&	2920.73	&	102.94	&	2255.61	&	108.35	&	$(+1/2)$	&	1875.04	&	118.28	&	1808.07	&	120.28	&	1766.02	&	21.76	\\
	&	1077.15	&	114.82	&	1116.75	&	116.27	&	1105.91	&	121.27	&		&	1079.70	&	131.17	&	1082.26	&	132.32	&	1072.08	&	134.34	\\
	&	617.39	&	175.83	&	616.55	&	182.48	&	614.06	&	189.51	&		&		&		&		&		&		&		\\
	&	589.44	&	192.13	&	592.50	&	194.22	&	596.38	&	199.86	&		&		&		&		&		&		&		\\
	&	419.55	&	27.71	&	418.78	&	27.40	&	417.25	&	26.66	&		&	404.65	&	6.03	&	404.65	&	6.82	&	404.29	&	7.99	\\
	&	212.32	&	-25.66	&	212.12	&	-27.38	&	212.32	&	-28.26	&		&	212.32	&	-31.54	&	212.32	&	-32.58	&	212.32	&	-32.52	\\
	&	202.59	&	-21.92	&	202.59	&	-22.97	&	202.68	&	-24.18	&		&	200.90	&	-26.58	&	200.99	&	-26.96	&	201.16	&	-27.34	\\
$(-1/2)$	&	14697.86	&	97.94	&	4032.16	&	102.43	&	2744.78	&	107.01	&	$(-1/2)$	&	1766.02	&	119.13	&	1719.37	&	120.76	&	1687.53	&	122.43	\\
	&	1077.15	&	114.82	&	1116.75	&	116.27	&	1105.91	&	121.27	&		&	1079.70	&	130.99	&	1082.26	&	132.32	&	1072.08	&	134.34	\\
	
	&	419.55	&	-4.56	&	418.40	&	-2.99	&	416.87	&	-0.42	&		&	404.65	&	25.76	&	404.65	&	25.14	&	404.29	&	25.28	\\
	&	212.32	&	-25.66	&	212.12	&	-27.38	&	212.12	&	-28.26	&		&	212.32	&	-31.54	&	212.32	&	-32.58	&	212.32	&	-32.52	\\
	&	199.05	&	-20.64	&	199.49	&	-21.48	&	199.75	&	-22.95	&		&	202.86	&	-27.35	&	202.86	&	-27.76	&	202.86	&	-28.18	\\
$(+3/2)$	&	1489.00	&	106.01	&	1484.15	&	109.22	&	1479.33	&	113.58	&	$(+3/2)$	&	1739.06	&	119.54	&	1719.37	&	120.76	&	1697.59	&	122.35	\\
	&	1095.27	&	113.95	&	1125.02	&	116.27	&	1116.75	&	121.17	&		&	1082.26	&	130.85	&	1087.43	&	132.17	&	1072.08	&	134.34	\\
	&	631.07	&	170.76	&	631.07	&	176.44	&	630.20	&	182.83	&		&	668.08	&	177.61	&	664.19	&	180.87	&		&		\\
	&	566.00	&	208.97	&	570.25	&	212.34	&	570.97	&	217.89	&		&	648.13	&	184.20	&	649.05	&	186.09	&		&		\\
	&	420.33	&	61.77	&	419.17	&	59.39	&	417.25	&	55.85	&		&	404.65	&	-12.78	&	404.65	&	-11.43	&	404.29	&	-8.59	\\
	&	212.32	&	-25.66	&	212.12	&	-27.38	&	212.12	&	-28.26	&		&	212.32	&	-31.54	&	212.32	&	-32.58	&	212.32	&	-32.52	\\
	&	205.61	&	-23.12	&	205.43	&	-24.00	&	205.06	&	-25.42	&		&	198.88	&	-25.54	&	199.14	&	-26.08	&	199.40	&	-26.58	\\
$(-3/2)$	&	2109.41	&	101.90	&	2007.20	&	105.43	&	1890.60	&	110.08	&	$(-3/2)$	&	1479.33	&	121.89	&	1474.54	&	123.15	&	1479.33	&	124.73	\\
	&	1084.84	&	114.38	&	1116.75	&	116.27	&	1114.02	&	121.23	&		&	1087.43	&	130.75	&	1092.65	&	132.01	&	1077.15	&	134.14	\\
	&	694.56	&	150.71	&	691.40	&	155.63	&	687.23	&	163.29	&		&	628.46	&	192.56	&	626.73	&	196.34	&	625.01	&	198.77	\\
	&	643.55	&	165.68	&	644.46	&	169.06	&	646.29	&	175.62	&		&	575.29	&	222.87	&	577.48	&	225.62	&	578.95	&	226.73	\\
	&	419.17	&	-35.11	&	418.01	&	-31.33	&	416.48	&	-26.87	&		&	405.01	&	45.89	&	404.65	&	43.96	&	404.29	&	41.92	\\
	&	212.32	&	-25.66	&	212.12	&	-27.38	&	212.12	&	-28.26	&		&	212.32	&	-31.54	&	212.32	&	-32.58	&	212.32	&	-32.52	\\
	&	195.22	&	-19.20	&	195.89	&	-20.45	&	196.82	&	-21.76	&		&	204.50	&	-28.12	&	204.32	&	-28.49	&	204.23	&	-28.82	\\
$(+1/2)$	&		&		&		&		&		&		&	$(+1/2)$	&	1455.70	&	122.30	&	1474.54	&	123.15	&	1493.88	&	124.25	\\
	&		&		&		&		&		&		&		&	1097.91	&	130.25	&	1097.91	&	131.69	&	1082.26	&	133.88	\\
	&	635.47	&	168.88	&	633.70	&	176.44	&	633.70	&	181.31	&		&	697.75	&	169.21	&	694.56	&	171.81	&	690.35	&	175.53	\\
	&	537.94	&	237.89	&	542.42	&	241.88	&	547.64	&	243.35	&		&	643.55	&	186.01	&	644.46	&	187.88	&	644.46	&	190.28	\\
	&	420.33	&	98.74	&	419.55	&	93.34	&	417.63	&	86.15	&		&	404.65	&	-32.05	&	404.65	&	-28.77	&	403.93	&	-24.62	\\
	&	212.32	&	-25.66	&	212.12	&	-27.38	&	212.12	&	-28.26	&		&	212.32	&	-31.54	&	212.32	&	-32.58	&	212.32	&	-32.52	\\
	&	208.91	&	-24.48	&	208.34	&	-25.49	&	207.86	&	-26.20	&		&	196.48	&	-24.61	&	196.99	&	-25.12	&	197.67	&	-25.82	\\
$(-1/2)$	&		&		&	1228.12	&	113.97	&	1279.87	&	116.64	&	$(-1/2)$	&	1262.14	&	125.78	&	1276.28	&	127.13	&	1287.10	&	127.99	\\
	&		&		&	1130.60	&	116.13	&	1127.81	&	120.74	&		&	1111.30	&	129.79	&	1111.30	&	131.27	&	1095.27	&	133.42	\\
	&	786.93	&	134.41	&	760.66	&	140.86	&	740.87	&	151.66	&		&	631.95	&	190.75	&	631.07	&	193.16	&	631.07	&	196.65	\\
	&	640.83	&	167.80	&	641.74	&	173.42	&	642.64	&	177.94	&		&	563.21	&	233.46	&	569.54	&	231.24	&	570.25	&	232.41	\\
	&	419.17	&	-62.67	&	418.01	&	-58.24	&	416.10	&	-51.20	&		&	405.01	&	67.44	&	404.65	&	64.11	&	404.65	&	60.55	\\
	&	212.32	&	-25.66	&	212.12	&	-27.38	&	212.12	&	-28.26	&		&	212.32	&	-31.54	&	212.32	&	-32.58	&	212.32	&	-32.52	\\
	&	189.06	&	-17.17	&	190.56	&	-18.38	&	192.49	&	-20.11	&		&	206.82	&	-29.23	&	206.45	&	-29.52	&	205.98	&	-29.63	\\

\hline
\hline
\label{table:nonlin} 
\label{I}

\end{tabular}
\end{table}

\begin{table}[h]
\scriptsize
  \caption{Magic wavelengths (in nm)  of Sr$^+$ for different focusing angles 50$^\circ$, 60$^\circ$ and 70$^\circ$ of the LG beam for the transitions $5s_{1/2}(-1/2)  $  $\rightarrow$ $4d_{5/2}(m_J)  $.}
\centering
\begin{tabular}{cccccccccccccc}

\hline \hline
     \multicolumn{14}{c}{\textbf{Non-paraxial LG beam}}\\ 

   State  ($4d_{5/2}(m_J)$)    & $\lambda_{\textrm{magic}}^{50^\circ}$&$\alpha$ & $\lambda_{\textrm{magic}}^{60^\circ}$&$\alpha$&  $\lambda_{\textrm{magic}}^{70^\circ}$&$\alpha$& State     ($4d_{5/2}(m_J)$) & $\lambda_{\textrm{magic}}^{50^\circ}$&$\alpha$ & $\lambda_{\textrm{magic}}^{60^\circ}$&$\alpha$&  $\lambda_{\textrm{magic}}^{70^\circ}$&$\alpha$
           \\ [0.2ex]

 \hline 
 & \multicolumn{6}{c}{\textbf{OAM=+2, SAM=+1}} && \multicolumn{6}{c}{\textbf{OAM=+2, SAM=-1}}\\  
  \hline

$(+1/2)$	&	3927.88	&	100.17	&	2517.31	&	106.00	&	2052.40	&	112.18	&	$(+1/2)$	&	1852.17	&	118.93	&	1793.83	&	120.89	&	1739.06	&	122.92	\\
	&	1077.15	&	115.89	&	1103.23	&	119.70	&	1069.56	&	126.17	&		&	1087.43	&	131.29	&	1074.61	&	133.39	&	1079.70	&	135.08	\\
	&	617.39	&	177.85	&	614.89	&	186.71	&	612.41	&	196.04	&		&		&		&		&		&		&		\\
	&	590.96	&	193.35	&	594.05	&	198.04	&	600.31	&	202.71	&		&		&		&		&		&		&		\\
	&	419.55	&	27.04	&	418.01	&	26.84	&	415.72	&	26.52	&		&	404.29	&	5.89	&	404.29	&	6.90	&	404.29	&	8.39	\\
	&	212.22	&	-26.23	&	212.32	&	-27.71	&	212.22	&	-29.44	&		&	212.22	&	-32.29	&	212.52	&	-32.56	&	212.22	&	-33.14	\\
	&	202.59	&	-22.30	&	202.68	&	-23.71	&	202.77	&	-24.98	&		&	200.90	&	-26.66	&	201.07	&	-27.13	&	201.34	&	-27.69	\\
$(-1/2)$	&	6328.24	&	99.71	&	3120.78	&	105.14	&	2289.62	&	111.36	&	$(-1/2)$	&	1739.06	&	119.79	&	1700.13	&	121.56	&	1650.85	&	123.61	\\
	&	1077.15	&	116.05	&	1103.23	&	119.70	&	1069.56	&	126.17	&		&	1090.03	&	131.10	&	1077.15	&	133.27	&	1079.70	&	135.05	\\

	&	419.17	&	-4.40	&	417.63	&	-2.19	&	415.72	&	0.68	&		&	404.29	&	26.20	&	404.29	&	25.68	&	404.29	&	24.43	\\
	&	212.22	&	-26.23	&	212.32	&	-27.71	&	212.22	&	-29.44	&		&	212.22	&	-32.29	&	212.52	&	-32.56	&	212.22	&	-33.14	\\
	&	199.14	&	-20.73	&	199.66	&	-22.34	&	200.10	&	-23.84	&		&	202.86	&	-27.49	&	202.86	&	-27.92	&	202.86	&	-28.32	\\
$(+3/2)$	&	1484.15	&	107.39	&	1479.33	&	111.66	&	1479.33	&	116.62	&	$(+3/2)$	&	1732.45	&	119.83	&	1706.49	&	121.50	&	1681.30	&	123.42	\\
	&	1095.27	&	115.33	&	1114.02	&	119.05	&	1082.26	&	125.77	&		&	1092.65	&	130.97	&	1077.15	&	133.19	&	1082.26	&	134.95	\\
	&	631.07	&	172.27	&	630.20	&	179.39	&	629.33	&	188.18	&		&	667.11	&	179.06	&	660.34	&	183.18	&		&		\\
	&	567.41	&	209.40	&	570.25	&	214.61	&	572.40	&	222.72	&		&	648.13	&	184.71	&	649.05	&	186.68	&		&		\\
	&	419.94	&	61.30	&	418.40	&	57.72	&	416.10	&	53.64	&		&	404.29	&	-12.25	&	404.29	&	-9.90	&	404.29	&	-7.10	\\
	&	212.22	&	-26.23	&	212.32	&	-27.71	&	212.22	&	-29.44	&		&	212.22	&	-32.29	&	212.52	&	-32.56	&	212.22	&	-33.14	\\
	&	205.52	&	-23.18	&	205.24	&	-24.35	&	204.87	&	-25.69	&		&	198.97	&	-25.73	&	199.31	&	-26.30	&	199.66	&	-26.87	\\
$(-3/2)$	&	2071.06	&	103.32	&	1947.15	&	107.98	&	1815.27	&	113.42	&	$(-3/2)$	&	1474.54	&	122.38	&	1479.33	&	123.94	&	1479.33	&	125.58	\\
	&	1087.43	&	115.55	&	1111.30	&	119.44	&	1074.61	&	126.06	&		&	1097.91	&	130.71	&	1082.26	&	132.99	&	1084.84	&	134.74	\\
	&	694.56	&	152.80	&	689.31	&	159.51	&	681.07	&	169.57	&		&	627.59	&	193.70	&	626.73	&	196.76	&	625.01	&	200.22	\\
	&	644.46	&	167.38	&	646.29	&	173.59	&	646.29	&	180.33	&		&	576.02	&	223.86	&	577.48	&	226.28	&	579.69	&	227.60	\\
	&	418.78	&	-34.39	&	417.25	&	-29.38	&	415.35	&	-23.95	&		&	404.29	&	45.43	&	404.65	&	43.94	&	404.65	&	41.00	\\
	&	212.22	&	-26.23	&	212.32	&	-27.71	&	212.22	&	-29.44	&		&	212.22	&	-32.29	&	212.52	&	-32.56	&	212.22	&	-33.14	\\
	&	195.30	&	-19.68	&	196.31	&	-21.02	&	197.50	&	-22.74	&		&	204.41	&	-28.21	&	204.32	&	-28.63	&	204.14	&	-29.02	\\
$(+1/2)$	&		&		&		&		&	1165.30	&	123.12	&	$(+1/2)$	&	1460.36	&	122.73	&	1484.15	&	123.80	&	1498.79	&	125.28	\\
	&		&		&		&		&	1136.24	&	123.94	&		&	1105.91	&	130.59	&	1090.03	&	132.75	&	1090.03	&	134.53	\\
	&	635.47	&	171.08	&	633.70	&	178.77	&	632.82	&	186.03	&		&	695.62	&	170.93	&	691.40	&	174.00	&	686.20	&	178.48	\\
	&	539.21	&	239.06	&	545.02	&	242.09	&	552.28	&	243.70	&		&	643.55	&	187.15	&	644.46	&	189.10	&	645.37	&	191.67	\\
	&	420.33	&	97.01	&	418.78	&	90.44	&	416.48	&	80.76	&		&	403.93	&	-30.40	&	403.93	&	-26.88	&	404.29	&	-22.59	\\
	&	212.22	&	-26.23	&	212.32	&	-27.71	&	212.22	&	-29.44	&		&	212.22	&	-32.29	&	212.52	&	-32.56	&	212.22	&	-33.14	\\
	&	208.72	&	-24.75	&	208.15	&	-26.05	&	207.58	&	-27.21	&		&	196.56	&	-24.70	&	197.33	&	-25.37	&	198.02	&	-26.09	\\
$(-1/2)$	&	1168.29	&	113.24	&	1262.14	&	115.47	&	1328.38	&	119.09	&	$(-1/2)$	&	1265.65	&	126.00	&	1279.87	&	127.27	&	1294.41	&	128.65	\\
	&	1136.24	&	113.69	&	1127.81	&	118.73	&	1103.23	&	124.89	&		&	1116.75	&	130.07	&	1100.56	&	132.28	&	1100.56	&	134.26	\\
	&	776.21	&	137.65	&	750.63	&	146.66	&	725.53	&	158.38	&		&	631.95	&	191.87	&	631.95	&	194.58	&	630.20	&	197.79	\\
	&	640.83	&	168.57	&	641.74	&	174.84	&	642.64	&	182.21	&		&	565.30	&	233.57	&	569.54	&	231.96	&	570.97	&	234.22	\\
	&	418.40	&	-61.56	&	417.25	&	-55.41	&	414.97	&	-46.65	&		&	404.29	&	66.53	&	404.65	&	62.78	&	404.65	&	57.52	\\
	&	212.22	&	-26.23	&	212.32	&	-27.71	&	212.22	&	-29.44	&		&	212.22	&	-32.29	&	212.52	&	-32.56	&	212.22	&	-33.14	\\
	&	189.45	&	-17.39	&	191.44	&	-19.02	&	193.80	&	-21.77	&		&	206.73	&	-29.36	&	206.17	&	-29.50	&	205.70	&	-29.76	\\

\hline
\hline
\label{table:nonlin} 
\label{I}

\end{tabular}
\end{table}




\end{document}